%% file: paper.tex
\documentclass[sigconf]{acmart}

\usepackage{lipsum}
\usepackage{multirow}
\usepackage{graphicx}
\usepackage{booktabs}
\usepackage{makecell} 
\usepackage{stfloats}
\usepackage{longtable}

\newcommand{\cmt}[1]{}

\usepackage{soul}
\usepackage{color}
\newcommand{\hlrev}[1]{{\sethlcolor{yellow}\hl{#1}}}

\AtBeginDocument{%
  \providecommand\BibTeX{{%
    \normalfont B\kern-0.5em{\scshape i\kern-0.25em b}\kern-0.8em\TeX}}}

\setcopyright{acmcopyright}
\copyrightyear{2024}
\acmYear{2024}
\acmDOI{10.1145/3613905.3636287}

\acmConference[CHI EA '24]{Extended Abstracts of the CHI Conference on Human Factors in Computing Systems}{May 11--16,
  2024}{Honolulu, HI, USA}
\acmISBN{979-8-4007-0331-7/24/05}

\begin{document}

\title{CUI@CHI 2024: Building Trust in CUIs — From Design to Deployment}

\author{Smit Desai}
\email{smitad2@illinois.edu}
\affiliation{%
  \institution{University of Illinois at Urbana-Champaign}
  \city{Urbana-Champaign}
  \country{USA}
}

\author{Christina Wei}
\email{christina.wei@mail.utoronto.ca}
\orcid{0000-0001-6046-7938}
\affiliation{%
  \institution{University of Toronto}
  \city{Toronto}
  \country{Canada}
}

\author{Jaisie Sin}
\email{jaisie.sin@ubc.ca}
\affiliation{%
  \institution{University of British Columbia}
  \city{Toronto}
  \country{Canada}
}

\author{Mateusz Dubiel}
\email{mateusz.dubiel@uni.lu}
\orcid{0000-0001-8250-3370}
\affiliation{%
  \institution{University of Luxembourg}
  \city{Esch-sur-Alzette}
  \country{Luxembourg}
}

\author{Nima Zargham}
\email{zargham@uni-bremen.de}
\orcid{0000-0003-4116-0601}
\affiliation{
  \institution{Digital Media Lab, University of Bremen}
  \city{Bremen}
  \country{Germany}
}

\author{Shashank Ahire}
\email{shashank.ahire@hci.uni-hannover.de }
\affiliation{%
  \institution{Leibniz University}
  \city{Hannover}
  \country{Germany}
}

\author{Martin Porcheron}
\email{martin@boldinsight.uk}
\orcid{0000-0003-3814-7174}
\affiliation{
  \institution{Bold Insight}
  \city{London}
  \country{UK}
}

\author{Anastasia Kuzminkyh}
\email{anastasia.kuzminykh@utoronto.ca}
\affiliation{%
  \institution{University of Toronto}
  \city{Toronto}
  \country{Canada}
}

\author{Minha Lee}
\email{m.lee@tue.nl}
\affiliation{
  \institution{Eindhoven University of Technology}
  \city{Eindhoven}
  \country{The Netherlands}
}

\author{Heloisa Candello}
\email{heloisacandello@br.ibm.com}
\affiliation{
  \institution{IBM Research}
  \city{São Paulo}
  \country{Brazil}
}

\author{Joel Fischer}
\email{Joel.Fischer@nottingham.ac.uk}
\affiliation{
  \institution{University of Nottingham}
  \city{Nottingham}
  \country{UK}
}

\author{Cosmin Munteanu}
\email{cosmin@taglab.ca}
\affiliation{%
  \institution{University of Waterloo}
  \city{Waterloo}
  \country{Canada}
}

\author{Benjamin R Cowan}
\email{benjamin.cowan@ucd.ie}
\affiliation{
  \institution{University College Dublin}
  \city{Dublin}
  \country{Ireland}
}

\renewcommand{\shortauthors}{Desai et al.} 

\begin{abstract}
Conversational user interfaces (CUIs) have become an everyday technology for people the world over, as well as a booming area of research. Advances in voice synthesis and the emergence of chatbots powered by large language models (LLMs), notably ChatGPT, have pushed CUIs to the forefront of human-computer interaction (HCI) research and practice. Now that these technologies enable an elemental level of usability and user experience (UX), we must turn our attention to higher-order human factors: trust and reliance. In this workshop, we aim to bring together a multidisciplinary group of researchers and practitioners invested in the next phase of CUI design. Through keynotes, presentations, and breakout sessions, we will share our knowledge, identify cutting-edge resources, and fortify an international network of CUI scholars. In particular, we will engage with the complexity of trust and reliance as attitudes and behaviours that emerge when people interact with conversational agents.
\end{abstract}




\maketitle

\input{sections/1-motivation}
\input{sections/2-organizers}
\input{sections/4-inpeson-hybrid}
\input{sections/5-asynchronous-engagement}
\input{sections/6-workshop-activities}
\input{sections/7-post-workshop-plans}
\input{sections/3-plans-to-publish}

\input{sections/8-call-for-participation}
\begin{acks}
We thank Dr. S. Shyam Sundar for accepting the invitation to present the keynote address at the workshop. We also thank Dr. Katie Seaborn for contributing to this workshop proposal.
\end{acks}

\bibliographystyle{config/ACM-Reference-Format}
\bibliography{references}

\appendix

\end{document}

%% file: sections/1-motivation.tex
\section{Motivation}
Conversational User Interfaces (CUIs), such as virtual assistants like Apple Siri and Amazon Alexa, alongside Large Language Models (LLMs) like OpenAI's ChatGPT and Google Bard, are steadily becoming integral to our daily routines. Users engage in a wide range of activities with these technologies, from setting reminders and shopping online to seeking information and having casual conversations \cite{Sciuto_Saini_Forlizzi_Hong_2018}. Regardless of the type of interaction—transactional or social—it necessitates the establishment of trust \cite{Rheu_Shin_Peng_Huh-Yoo_2021}. This trust is paramount because it not only ensures the accuracy and reliability of information and services but also influences how comfortable users feel when sharing personal data or forging deep connections with these AI-powered systems \cite{Dubiel_Daronnat_Leiva_2022}. In this workshop, our main goal is to unite dedicated researchers from the CHI community for an in-depth exploration of CUIs from a trust perspective. We will delve into the unique nuances of trust within CUIs, addressing open questions and charting the course for future developments in this evolving landscape. 

Despite trust being crucial in the adoption and continued use of automated systems \cite{Rheu_Shin_Peng_Huh-Yoo_2021, sin2021vui, pradhan2020use, muir1987trust}, it is understudied in CUI, as exemplified by calls for research in CUI’19 \cite{Edwards_Sanoubari_2019} and again in CUI’22 \cite{Dubiel_Daronnat_Leiva_2022}. This lack of research is perhaps most evident in the opacity associated with the fundamental nature of trust, and specifically its conceptualization \cite{law2022conversational}. While the term `trust' finds frequent usage within HCI \cite{Bach_Khan_Hallock_Beltrão_Sousa_2022} and the more specialized literature on CUIs \cite{Edwards_Sanoubari_2019, Dubiel_Daronnat_Leiva_2022, Wei_Kim_Kuzminykh_2023, lee2021brokerbot}, a clear and generally accepted definition within this context remains elusive. In CUIs, trust can manifest in diverse ways: (1) it may encompass both cognitive elements rooted in thinking and judgment and affective aspects rooted in feelings \cite{Ueno_Sawa_Kim_Urakami_Oura_Seaborn_2022}, (2) it often comprises attitudinal \cite{Lee_See_2004} and behavioral components \cite{Söllner_Hoffmann_Leimeister_2016}, (3) it can be user-centric or agent-centric (including the owner of the agent) \cite{Ueno_Sawa_Kim_Urakami_Oura_Seaborn_2022}, and (4) its nature is inherently context-dependent \cite{Bach_Khan_Hallock_Beltrão_Sousa_2022, Weitz_Schiller_Schlagowski_Huber_André_2019}. This complexity is particularly pronounced due to the natural modality employed by users when interacting with these systems, which introduces a unique set of trust dynamics. Given the `black-box' nature of conversational interfaces, users frequently rely on anthropic tendencies \cite{Liao_Vaughan_2023}. Anthropomorphism only serves as a means of sense-making or constructing a mental model in this context \cite{Desai_Twidale_2022}. However, it is noteworthy that anthropomorphism is often intentionally incorporated into these interfaces, including the utilization of pronouns, natural-sounding voices, and the emulation of human-human communication models, among other techniques \cite{Kontogiorgos_Pereira_Andersson_Koivisto_Gonzalez_Rabal_Vartiainen_Gustafson_2019, Pradhan_Findlater_Lazar_2019, Desai_Twidale_2022, Kuzminykh_Sun_Govindaraju_Avery_Lank_2020, Strathmann_Szczuka_Krämer_2020}. 

Overtrust and undertrust in CUIs can have detrimental consequences \cite{muir1987trust, skitka1999does}, potentially resulting in discontinued or infrequent use of such systems or conversely, fostering addiction and diminishing user agency. \cite{Luger_Sellen_2016, Cowan_Pantidi_Coyle_Morrissey_Clarke_Al-Shehri_Earley_Bandeira_2017, Cho_Lee_Lee_2019}. Specifically, in critical domains such as finance and health, overreliance on voice-based advice may lead to misguided financial decisions \cite{Dubiel_Daronnat_Leiva_2022} or, in the case of health, the inadvertent disclosure of sensitive medical information without a complete understanding of the associated privacy risks, including potential breaches of confidentiality and unauthorized access to personal health data  \cite{Cho_2019}. These concerns may be exacerbated by demographic factors, such as age, where older individuals may be more vulnerable to trust-related issues due to their potentially limited familiarity with AI technologies \cite{Pradhan_Findlater_Lazar_2019}. Moreover, these challenges can vary across cultural contexts, as highlighted in the Global South, where cultural factors and technological savviness can significantly influence the levels of trust individuals place in these technologies \cite{Okolo_Kamath_Dell_Vashistha_2021}. Furthermore, as the conversational abilities of these technologies advance, the emotional bonds they develop with users may deepen, intensifying the emotional distress experienced in the event of dysfunction or breakdown \cite{Lee_Frank_De_Kort_IJsselsteijn_2022}, further underscoring the need for careful consideration of trust dynamics in CUIs.

Within the landscape of trust calibration, diverse strategies have emerged, ranging from efforts to introduce a sense of `distrust' \cite{Pinhanez_2021} to persistent calls for transparency, notably advocated by the burgeoning field of Explainable AI (XAI) \cite{Weitz_Schiller_Schlagowski_Huber_André_2021}. However, a comprehensive discussion of trust requires a multifaceted examination from various perspectives, encompassing specialized areas within CHI, such as Human-AI Interaction (HAII), Human-Robot Interaction (HRI), XAI, and CUI. Our objective is to foster collaborations that could help to converge these diverse approaches and ultimately lead to the development of a cohesive framework for a systematic study of trust. In our workshop discussions, we will reflect on: 

\begin{itemize}
\item \textbf{Conceptual Clarity of Trust in CUIs:} The foundation of our exploration involves refining and defining the conceptual boundaries of trust within the context of CUIs. We aim to foster a more informed and unified discourse by establishing a clear and shared understanding of trust.

\item \textbf{Design Techniques and Trust Calibration:} We will scrutinize design strategies that can inadvertently result in the miscalibration of user trust and raise ethical concerns. An illustrative example is the integration of human-like traits into CUIs, which can impact user perceptions.

\item \textbf{Interconnections of Trust with Other Values:} Trust can exhibit a delicate balance, be diminished, or even conflict with other values such as privacy, transparency, or usability. This exploration discusses the emergence of tensions in specific scenarios and highlights how design decisions in CUIs can significantly impact these perceptions.

\item \textbf{Integrative Measurement Methods:} Developing comprehensive methods for measuring trust, combining objective and subjective scales, is important to our inquiry to mitigate overtrust and undertrust.

\item \textbf{Temporal evolution of trust:} Trust is a dynamic concept that evolves over time and context. We will explore the temporal aspects of trust, examining how it fluctuates in different stages of interaction and adoption.
\end{itemize}

By exploring these and other emergent topics, we aspire to foster an environment where interdisciplinary discussions on trust can flourish, drawing inspiration from diverse strategies within trust calibration and design. In alignment with the overarching theme of the ACM CUI'24 conference, which focuses on designing trustworthy conversational interfaces, our workshop aims to contribute a deeper understanding of trust in CUIs across all stages—from design to deployment.


%% file: sections/2-organizers.tex
\section{Organizers}
This workshop is being organized by a team of experienced researchers who bring expertise from diverse fields aligned with our objectives. Collectively, the organizers have a track record of publishing on a wide range of topics, including conversational user interfaces~\cite{ zargham2021MultiAgent, zargham2022Proactive,dubiel2020persuasive, dubiel2019inquisitive, lee2021brokerbot, Lee_Frank_De_Kort_IJsselsteijn_2022, clark2019makes}, human-robot interaction~\cite{McMillan2023,boudouraki2023being}, multi-modal communication~\cite{bonfert2021evaluation, Avanesi2023Multimodal}, and privacy and security~\cite{Bahrini2020GoodEvil, Bahrini2022Long, desai2023using, law2022conversational}. Integrating these disciplines shall facilitate collaboration among our trans-disciplinary participants and bridge the gap between these research areas.

\textbf{Smit Desai} is a Ph.D. candidate in the School of Information Sciences at the University of Illinois, Urbana-Champaign. His primary research focus centers around comprehending the mental models of users as they engage with conversational agents, utilizing innovative research techniques such as metaphor analysis \cite{Smit2023Metaphors}. He leverages this valuable insight to advance the development of conversational agents in diverse social roles, including educators \cite{SmitCHI2022} and storytellers \cite{SmitCUI}. Smit is a Provocation Papers Co-Chair at the upcoming ACM CUI 2024 conference. His research has yielded publications in esteemed HCI forums like CHI, TOCHI, CSCW, and CUI.

\textbf{Christina Wei} is a Ph.D. student in the School of Information at the University of Toronto, Canada. Her research focuses on designing conversation architecture for AI agents \cite{Wei_Kim_Kuzminykh_2023} in human-AI collaborative decision-making scenarios. She has previously organized a workshop at ACM CUI 2023, and she is on the organizing committee for the upcoming ACM CUI 2024 conference.

\textbf{Jaisie Sin} is a Post-Doctoral Fellow at the Okanagan Visualization \& Interaction Lab at the University of British Columbia. Her research focuses on the inclusive design of conversational interfaces for underrepresented users, with a primary focus on ageing. She is an Accessibility Chair at the upcoming CUI 2024 conference. She has also been a Full Papers Co-Chair at CUI '23, and co-organizer of previous CUI conferences related workshops at CHI '19–'23 (lead in '23), IUI '20–'21, and CSCW '20.

\textbf{Mateusz Dubiel} is a Research Associate in the Department of Computer Science at the University of Luxembourg, where he works on development and evaluation of conversational agents. Specifically, his current research focuses on assessment of cognitive and usability implications of interfaces that feature speech, and exploration of their potential to persuade and inspire positive behavioural change in users. He served as Short Papers Chair for CUI '22 and will be one of General Chairs for the upcoming CUI '24.

\textbf{Nima Zargham} is a Ph.D. student in the Digital Media Lab at the University of Bremen. His research focuses on human-centered approaches for designing speech-based systems that elicit desirable user experiences. Nima has previously organized CUI-related workshops at notable conferences such as ACM/IEEE HRI 2023 and ACM CUI 2023. Additionally, he served as a local chair at the ACM CHI-PLAY 2022 conference. His research efforts have resulted in publications featured in prestigious HCI venues, including CHI, CUI, and CHI-PLAY.

\textbf{Shashank Ahire} is a PhD candidate in Mensch Computer Interaktion group at Leibniz University Hannover. He is interested in developing CUIs to improve the health and well-being of knowledge Workers while at work \cite{ahire-UWA}. Previously, he has also researched on designing and developing CUIs for marginalized populations in the global south \cite{ahire-human-plus-AI}.

\textbf{Martin Porcheron} is a Senior UX Researcher at Bold Insight UK, based in London. His recent work focuses on the user experience of interactive “AI” systems—from voice interfaces through to robot interfaces. In his Ph.D., he also studied public places like cafés and pubs, to understand how we use devices like smartphones while we are socialising. As well as organising a number of workshops at international conferences, he was Full Papers Chair for the inaugural CUI '19 conference and Programme Chair for CUI '20 and '21, and was General Chair for CUI '23.

\textbf{Anastasia Kuzminkyh} is an Assistant Professor in Human-Computer Interaction at the University of Toronto in the Faculty of Information and the Director of the Toronto Human-AI Interaction Research School (THAI RS). Her work touches on diverse technologies, including conversational agents, large language models, and AI-based decision support systems, exploring the mechanisms driving the perceptions of different types of AI systems, such as trust, anthropomorphization, or perceived reliability to inform the system and algorithm design requirements for effective, efficient, and ethical AI.

\textbf{Minha Lee} is an assistant professor at the Eindhoven University of Technology in the Department of Industrial Design, with a background in philosophy, digital arts, and HCI. Her research concerns morally relevant interactions with various agents like robots or chatbots, exploring moral concepts like compassion and trust. She organized workshops such as CHI 2022 workshop on CUI Ethics, CSCW 2022 workshop on Activism in Academia, HRI 2021 workshop on Artificial Identity and more, as well as serving as a General Chair for CUI '23, Full papers for Chair CUI '21, and Provocation papers Chair for CUI '20,

\textbf{Heloisa Candello} is a research scientist and a manager of the Human-centered and  Responsible technologies group at IBM Research. Her work focuses on human and social aspects of Artificial Intelligence systems, particularly conversational user interfaces. Currently, Heloisa is leading a project that aims to bring “conscious” access to micro-credit by enhancing non-traditional financial practices of low-income small business owners with AI technology in the Global South. Her research resulted in several publications in leading conferences (CHI, CUI, CSCW, DRS) and recognition in the HCI and Design field.

\textbf{Joel Fischer} is Professor of Human-Computer Interaction at the University of Nottingham, and Research Director of two national programmes on Responsible and Trustworthy AI (TAS and RAI UK). Joel's interests include close studies of conversational interaction by adopting a conversation analytic perspective. He has published widely and co-organised many workshops on the topic at CHI, CUI, CSCW, and HRI.

\textbf{Cosmin Munteanu} is an Associate Professor and Schlegel Research Chair in Technology for Healthy Aging at the Department of Systems Design Engineering, University of Waterloo, and Director of the Technologies for Ageing Gracefully lab. Cosmin takes a primarily ethnomethodological approach to study how voice- and language-enabled interfaces should be designed in a safe, effective, inclusive, and ethical manner, in order to empower digitally underrepresented groups such as older adults. Cosmin is the co-founder of the new ACM Conversational User Interfaces conference series and one of the early researchers to bring speech processing and interface design research together. He is also serving as a member of the ACM SIGCHI Ethics Committee where he contributes perspectives on ethical research with underrepresented user groups.

\textbf{Benjamin Cowan} is a Professor at  University College Dublin's School of Information and Communication Studies. His research focuses on understanding user interaction with speech-based conversational user interfaces. He is co-founder of the ACM Conversational User Interfaces conference (ACM CUI) and has published extensively on user interaction issues with speech interfaces within the HCI community. He is an Associate Editor of the International Journal of Human-Computer Studies, has served as CHI AC, and was one of the inaugural SCs of the Understanding People: Quantitative Methods subcommittee at CHI 2023.

%% file: sections/4-inpeson-hybrid.tex
\section{In-person and Hybrid}
Based on our past experience in conducting workshops, we are aware of the various limitations attendees may face, (visa issues, financial constraints, and personal commitments) which can prevent them from attending in-person workshops. Therefore, we plan to host both an in-person and a hybrid workshop. Similar to the previous year, we aim to facilitate synchronous discussions for in-person attendees (with dedicated cameras) and online participants. Additionally, for the purpose of brainstorming and idea generation, we intend to synchronize the experiences of both in-person and online attendees by utilizing online platforms such as Miro.

We are committed to supporting the CHI2024 initiative aimed at creating an inclusive environment for in-person and hydrid attendees\footnote{\url{https://chi2024.acm.org/for-attendees/accessibility-faq/}}. In line with this commitment, we will kindly request all interested attendees to follow the `ACM Guide for Accessible Submission\footnote{\url{https://sigchi.org/conferences/author-resources/accessibility-guide/}}' when preparing their submissions. Additionally, we will encourage them to conduct an accessibility check\footnote{\url{https://www.youtube.com/watch?v=G8EYDJ56c6k}} before finalizing their submissions. To ensure an inclusive experience during the workshop, we will provide automatic transcription of keynote, and paper presentations for online participants.


%% file: sections/5-asynchronous-engagement.tex
\section{Asynchronous Engagement}


To advertise our workshop, we will use our existing channels: email list (cui-announcements@listserv.acm.org), Twitter handle (\url{https://twitter.com/ACM_CUI}), Discord channel (\url{https://discord.gg/rDmctrVb}) and workshop website (\url{ https://cui.acm.org/workshops/CHI2024/}). During the workshop, we will create a sub-channel (CUI@CHI2024) on existing Discord to disseminate information.


We will encourage participants to create 3-minute video presentations, which will be archived on the CUI channel on YouTube. The YouTube video of the presentation will offer automated subtitles to ensure accessibility for the uploaded presentations. We will record the keynote presentation and a summary of the discussions during the workshop. The video recordings will be accessible on Discord for participants to view asynchronously.




%% file: sections/6-workshop-activities.tex
\section{Workshop Activities}

We plan for a one-day, in-person/hybrid workshop structured in a series of presentations, activities, and structured discussions to exchange insights and lessons learned related to the building trust in CUIs from design to deployment. We are aiming for approximately 20 participants, which has been the norm in the past CUI workshops. Our tentative schedule is below.

\begin{enumerate}
    \item \textbf{Introduction (15 mins):} We start with brief introductions from organizers, highlighting the goals of the workshop to build trustworthy CUIs, and go over plans for the day.
    \item \textbf{Keynote Address (45 mins):} Our keynote address will be delivered by \textbf{Dr. S. Shyam Sundar}. He is the founder of the Media Effects Research Laboratory, a leading facility in the USA. He also serves as the Director of the Center for Socially Responsible Artificial Intelligence, an interdisciplinary consortium at Penn State University, USA dedicated to AI development with a strong emphasis on its social and ethical implications, aiming to enhance societal good and mitigate misuse. The speech will kick off the workshop to discuss different aspects of building trust with machines, such as explainable AI, trust calibration, and specific considerations for conversational interfaces. This keynote will set a solid foundation for the upcoming breakout discussions.
    \item \textbf{Introduction \& Presentations (60 mins):} Each participant will have 3-5 minutes to introduce themselves and present their accepted paper.
    \item \textbf{Break (15 mins)}
    \item \textbf{Breakout Session 1 - Prototype activity (60 mins):} Participants  will be divided into 3 or 4 groups for small group discussions, with a designated facilitator assigned for each group. Each group will be given a scenario to design a trustworthy CUI interaction. For example, designing a CUI to provide financial guidance to help individual stay within their budgets, or designing a CUI as a social companion for older adults. At the end of the session, each group will present their CUI prototype and discuss the  elements they selected to incorporate in their design, as well as received feedback on their design from the broader audience.
    \item \textbf{Lunch (60 mins)}
    \item \textbf{Breakout Session 2 - Design Techniques (60 mins):} Participants and facilitators will go back to their small groups to discuss different design techniques for building trust in CUIs related to their given scenario. This includes questions such as: What are the different trust design techniques for CUIs? What are the contextual considerations when applying these design techniques? What improvements can be made to the prototype in Breakout 1 based on these design techniques?  A sharing session will follow for each group to present a summary of their discussions as well as proposed design guidelines.
    \item \textbf{Breakout Session 3 - Evaluation Techniques \& Methods (60 mins):} The session will be introduced by Dr. Katie Seaborn. They are an Associate Professor at Tokyo Institute of Technology and founder of the Aspirational Computing Lab. Their research focuses on voice UX, AI, and intersectional design. Afterwards, participants and facilitators will go back to their small groups to discuss methods for measuring trust related to their given scenario. This includes questions such as: What are the existing methods used to measure trust with CUIs? What aspects of trust do these methods measure? What are some additional considerations for measuring trust that need to be taken into account for the given scenario?  A sharing session will follow for each group to present a summary of their discussions.
    \item \textbf{Break (15 mins)}
    \item \textbf{Closing (60 mins):} The organizers will end the workshop by synthesizing the main points from previous sessions. Participants and organizers will reflect together to develop tangible research topics and potential outcomes for CUI research. This will help in the development of future directions for the CUI community to consider in relation to building trustworthy CUIs, including how to plan for upcoming CUI conference series and workshops by taking into account participants’ views. 
    
\end{enumerate}

%% file: sections/7-post-workshop-plans.tex
\section{Post-Workshop Plans}

The expected workshop outcomes include:
\begin{itemize}
\item Reconnecting, sustaining, and extending the existing CUI community of researchers.
\item Invitation of participants to trustworthy design initiatives within the CUI community. 
\item Identification of major themes in trustworthy design relevant to CUI research.
\item A proposal for a special issue at relevant journals to highlight trust building design considerations relevant to CUI research, e.g., ACM interactions magazine, International Journal of Human-Computer Interaction, International Journal of Human-Computer Studies. 
\item A white paper summarizing outcomes from the workshop to be posted on our workshop's website, as well as socialized through platforms such as Twitter, Discord, and Medium. 
\item Invitation for a selection of papers to be published in workshop proceedings as outlined below.
\end{itemize}

%% file: sections/3-plans-to-publish.tex
\section{Plans to Publish Workshop Proceedings}

The primary objective of this workshop is to foster collaboration among individuals with diverse backgrounds and perspectives, all of whom share an interest in examining the role of trust in interactions with CUIs and strategies for establishing trust between users and these agents. 
All accepted contributions from participants will be published on our workshop's website to ensure the broad dissemination of the workshop's findings. 
Additionally, a selection of the accepted papers will be included in the workshop proceedings, which we aspire to publish on \textit{arXiv} \footnote{https://arxiv.org/} or \textit{ceur-ws} \footnote{https://ceur-ws.org/}.
Ideas and discussion points will be documented throughout the session on an open online platform, both during and after the symposium. Workshop participants will also be invited to engage in future collaborative projects that emerge from the discussions and talks held during the workshop. 
Finally, we will strongly encourage all participants to submit extended versions of their contributions for consideration in a special journal issue, which we are committed to facilitating and making a reality.

%% file: sections/8-call-for-participation.tex
\section{Call for Participation}

How can we advance research on Conversational User Interfaces (CUIs) by including consideration of trust? Since CUIs, such as Apple Siri, Amazon Alexa and Google Home are becoming increasingly used in privacy-sensitive contexts such as self-reflection, tutoring or therapy, investigation of trust and impact of user’s reliance and agency becomes ever more pertinent. What is more, proliferation of LLMs, such as OpenAI's ChatGPT and Google Bard and growing role their role in assisting human decision-making also calls for investigation. 

Authors across all disciplines are invited to submit position papers to the CHI 2024 workshop on the Building Trust in CUIs—From Design to Deployment. The goal of this edition of the workshop is to encourage discussion on implementing ethical and user-oriented design practices to develop CUIs that are focused on enhancing user’s agency while preserving privacy and preventing overreliance. We expect that the workshop will encourage contributions and discussions from participants from both industry and academia. Papers can address the state of CUIs and their design for privacy-sensitive scenarios where trust plays an essential role, ways to practice design of trustworthy CUIs, and how we can move forward in design of CUIs that empower users and enhance their agency. 

Submissions are expected to be between 2 to 4 pages in the CHI Publication Format\footnote{https://chi2024.acm.org/submission-guides/chi-publication-formats/}, including references, authors’ work thus far, disciplinary background(s), as well as how they relate to the topics and goals of the workshop. We will select submissions based on quality; if accepted, papers will be featured on our website and at least one author must attend the workshop. Papers should be submitted to smitad2@illinois.edu. Those who cannot or do not wish to submit a position paper to the workshop can submit a one-page statement of interest for consideration. This should include why the applicant is interested in participating in the workshop, their relevancy to the workshop, and what the person expects to contribute.